\pdfoutput=1

\documentclass[11pt,twoside,a4paper,cmspaper,final,collab]{cms-tdr}
\def\svnVersion{43827M}\def\cmsCernNo{2011-016}\def\cmsCernDate{\today}\def\cmsMessage{Submitted to the Journal of High Energy Physics}

\begin{document}\cmsNoteHeader{SUS-10-007}

\hyphenation{had-ron-i-za-tion}
\hyphenation{cal-or-i-me-ter}
\hyphenation{de-vices}
\RCS$Revision: 43212 $
\RCS$HeadURL: svn+ssh://alverson@svn.cern.ch/reps/tdr2/papers/SUS-10-007/trunk/SUS-10-007.tex $
\RCS$Id: SUS-10-007.tex 43212 2011-03-02 15:42:32Z fronga $
\cmsNoteHeader{SUS-10-007} 
\title{Search for Physics Beyond the Standard Model in Opposite-sign Dilepton Events in pp Collisions at $\sqrt{s}=7$~TeV}

\address[neu]{Northeastern University}
\address[fnal]{Fermilab}
\address[cern]{CERN}
\author[cern]{The CMS Collaboration}

\newcommand{\mll}{\ensuremath{M_{\ell\ell}}}
\newcommand{\eepm}{\Pep\Pem}
\newcommand{\mmpm}{\Pgmp\Pgmm}
\newcommand{\empm}{\ensuremath{\Pe^\pm \Pgm^\mp}}
\newcommand{\ase}[2]{\ensuremath{_{~- #1}^{~+ #2}}}

\newcommand{\lumifinal}{34\pbinv}

\date{\today}

\abstract{
A search  is presented for physics  beyond the standard  model (SM) in
final states  with opposite-sign isolated lepton  pairs accompanied by
hadronic jets and missing  transverse energy.  The search is performed
using LHC  data recorded  with the CMS  detector, corresponding  to an
integrated luminosity of 34~pb$^{-1}$.  No evidence for an event yield
beyond  SM  expectations is  found.   An  upper  limit on  the  non-SM
contribution to the  signal region is deduced from  the results.  This
limit  is  interpreted  in  the  context of  the  constrained  minimal
supersymmetric  model.  Additional  information is  provided  to allow
testing the exclusion of specific models of physics beyond the SM.
}

\hypersetup{%
pdfauthor={CMS Collaboration},%
pdftitle={Search for Physics Beyond the Standard Model in Opposite-Sign Dilepton Events at sqrt(s) = 7 TeV},%
pdfsubject={CMS},%
pdfkeywords={CMS, physics, supersymmetry}}

\maketitle 

\section{Introduction}
\label{sec:intro}

In this paper we describe a search for physics beyond the standard model (BSM)
in a sample of proton-proton collisions at a centre-of-mass energy of 7~TeV.
The data sample was collected with the Compact Muon Solenoid (CMS) detector~\cite{JINST} at
the Large Hadron Collider (LHC) between March and
November of 2010 and corresponds to an integrated luminosity of \lumifinal.

The BSM signature in this search is motivated  by three general  considerations.
First, new particles predicted by BSM
physics scenarios are expected to be heavy, since they have so far eluded detection.
Second, BSM physics  signals with  high
enough  cross sections to  be observed  in our  current dataset are expected to be
produced  strongly,  resulting  in  significant hadronic  activity.
Third, astrophysical evidence for
dark matter  suggests~\cite{ref:DM,ref:DM2}  that the mass  of weakly-interacting
massive particles is of the  order of the electroweak symmetry breaking
scale. Such particles, if produced in pp collisions, could escape detection and give rise to
an apparent imbalance in the  event transverse energy. We therefore focus  on the
region  of  high missing transverse energy (\MET). An  example of  a  specific  BSM scenario  is
provided by R-parity conserving  supersymmetric (SUSY) models in which
new,  heavy  particles  are  pair-produced  and  subsequently  undergo
cascade       decays,      producing      hadronic       jets      and
leptons~\cite{Martin:fk,susy1,susy2,susy3,susy4,susy5,susy6}.
These cascade decays  may terminate  in the
production  of weakly-interacting massive  particles,  resulting in large \MET.

The results reported in this paper are part of a broad program of BSM searches
in events with jets and \MET, characterized by the number and
type of leptons in the final state.
Here we describe a search for events containing opposite-sign isolated
lepton pairs (\eepm, \empm, \mmpm) in addition to the jets
and \MET. Results from a complementary search with no electrons or muons in the
final state have already been reported in Ref.~\cite{ref:RA1}.

Our analysis strategy is as follows. In order to select dilepton events, we
use high-\pt\ lepton triggers and a preselection based
on that of the $\cPqt\cPaqt$ cross section measurement in the dilepton channel~\cite{ref:top}.
Good agreement is found between this
data sample and predictions from SM Monte Carlo (MC) simulations in terms of the event yields
and shapes of various kinematic distributions.
Because BSM physics is expected to have large hadronic activity and \MET\ as discussed
above, we  define a signal region
with requirements on these quantities to select about 1\%
of dilepton $\cPqt\cPaqt$ events, as predicted by MC.
The observed event yield in the signal region is compared with the predictions from two
independent background estimation techniques based on data control samples,
as well as with SM and BSM MC expectations.
Finally, the robustness of the result is confirmed by an independent analysis based on
hadronic activity triggers, different ``physics object'' reconstruction, and a complementary
background estimation method.

No specific BSM physics scenario, e.g.\ a particular SUSY model, has been used to optimize the search.
In order to illustrate the sensitivity of the search, a simplified and practical model of
SUSY breaking, the constrained minimal supersymmetric
extension of the standard model (CMSSM)~\cite{CMSSM,CMSSM2}, is used. The CMSSM is described by
five parameters: the universal scalar and gaugino mass parameters ($m_0$ and $m_{1/2}$, respectively),
the universal trilinear soft SUSY breaking parameter $A_0$, the
ratio of the vacuum expectation values of the two Higgs doublets ($\tan\beta$), and the sign of the
Higgs mixing parameter $\mu$. Throughout the paper, two CMSSM parameter sets, referred
to as LM0 and LM1~\cite{TDR}, are used to illustrate possible CMSSM yields. The parameter values
defining LM0 (LM1) are $m_0 = 200~(60) \GeVcc$, $m_{1/2} = 160~(250) \GeVcc$, $A_0 = -400~(0)\GeV$;
both LM0 and LM1 have $\tan\beta = 10$ and $\mu > 0$.
These two scenarios are beyond the exclusion reach
of previous searches performed at the Tevatron and LEP. They were recently excluded
by a search performed at CMS in events with jets and \MET~\cite{ref:RA1} based on the same data sample
used for this search. In this analysis, the LM0 and LM1 scenarios serve as benchmarks which
may be used to allow comparison of the sensitivity with other analyses.

\section{CMS Detector}

The central feature of the CMS apparatus is a superconducting
solenoid, 13~m in length and 6~m in diameter, which provides
an axial magnetic field of 3.8 T. Within the field volume are
several particle detection systems. 
Charged particle
trajectories are measured by silicon pixel and silicon strip trackers,
covering $0 \leq \phi \leq 2\pi$ in azimuth and $|\eta| < 2.5$ in pseudorapidity,
defined as $\eta = -\log [\tan \theta/2]$, where $\theta$ is the
polar angle of the trajectory of the particle with respect to
the counterclockwise proton beam direction. A crystal electromagnetic calorimeter
and a brass/scintillator hadronic calorimeter surround the
tracking volume, providing energy  measurements of electrons and
hadronic jets. Muons are identified and measured in gas-ionization detectors embedded in
the steel return yoke outside the solenoid. The detector is nearly
hermetic, allowing energy balance measurements in the plane
transverse to the beam direction. A two-tier trigger system selects
the most interesting pp collision events for use in physics analysis.
A more detailed description of the CMS detector can be found
elsewhere~\cite{JINST}.

\section{Event Selection}
\label{sec:eventSel}

Samples of  MC events are used to  guide the  design of  the analysis.
These      events     are      generated     using      either     the
\PYTHIA6.4.22~\cite{Pythia}  or  \MADGRAPH4.4.12~\cite{Madgraph} event
generators.    They   are   then   simulated  using   a   GEANT4-based
model~\cite{Geant} of the CMS  detector, and finally reconstructed and
analyzed using the same software as is used to process collision data.

We apply a  preselection   based  on  that of the $\cPqt\cPaqt$  cross section
measurement in  the dilepton channel~\cite{ref:top}.  Events
with     two     opposite-sign,     isolated    leptons     (\eepm,
$\empm$, or $\mmpm$) are selected. At least one of the leptons must
have  $\pt >  20\GeVc$ and  both must  have $\pt  > 10\GeVc$,  and the
electrons (muons) must have $|\eta| < 2.5$ ($|\eta| < 2.4$). In events
with more  than two such leptons,   the two  leptons with the
highest \pt are selected.  Events with an $\eepm$  or $\mmpm$ pair
with  invariant mass  between 76\GeVcc  and  106\GeVcc or  below
10\GeVcc are removed, in  order      to     suppress     Drell--Yan  (DY)
$\cPZ/\gamma^{*}\to\ell\ell$  events,  as  well  as   low  mass  dilepton
resonances.

Events are required to pass  at least one  of a set of  single-lepton or
double-lepton triggers.  The  efficiency for events containing two
leptons passing the analysis selection to  pass at  least  one of  these
triggers  is very  high, in  excess  of 99\%  for dilepton  $t\bar{t}$
events.

Because leptons produced in the  decays of low-mass particles, such as
hadrons containing $\cPqb$  and $\cPqc$ quarks,  are  nearly  always inside  jets,  they can  be
suppressed by requiring the leptons to be isolated in space from other
particles that carry a  substantial amount of transverse momentum. The
details   of   the  lepton   isolation   measurement   are  given   in
Ref.~\cite{ref:top}.   In  brief,   a cone is constructed     of  size
$\Delta{}R\equiv\sqrt{(\Delta\eta)^2+(\Delta\phi)^2}=0.3$  around  the
lepton  momentum  direction. The  lepton  relative  isolation is  then
quantified  by  summing the  transverse  energy  (as  measured in  the
calorimeters) and the transverse  momentum (as measured in the silicon
tracker) of  all objects  within this cone,  excluding the  lepton, and
dividing by  the lepton transverse momentum. The resulting quantity
is required to be  less than 0.15, rejecting
the large background arising from QCD production of jets.

We require the presence of at least two jets with  $\pt > 30\GeVc$ and  $|\eta| < 2.5$,
separated  by $\Delta  R  >$  0.4 from  leptons  passing the  analysis
selection   with   $\pt  >   10\GeVc$.    The  anti-$k_T$   clustering
algorithm~\cite{antikt}  with  $\Delta{}R  =  0.5$  is  used  for  jet
clustering.  Jets are  reconstructed using calorimeter information and
their energies  are corrected using  reconstructed tracks~\cite{jets}.
The event is required to satisfy $\HT > 100\GeV$, where \HT\ is defined as
the scalar  sum of the transverse  energies of the  selected jets.  In
addition,  the  \MET\ in  the event is required to exceed  50\GeV. Several
techniques are used in CMS for calculating \MET~\cite{MET}.  Here, the
raw \MET, calculated from  calorimeter signals in the range $|\eta| < 5.0$,
is corrected by taking into  account   the  contributions  from   minimally  interacting
muons. The  \MET\ is further  corrected on a track-by-track  basis for
the  expected response  of  the calorimeter  derived from  simulation,
resulting in  an improved \MET\ resolution.

The data yields and corresponding MC predictions after this event preselection
are given in Table~\ref{tab:yields}. The MC yields are normalized to~\lumifinal\ using
next-to-leading order (NLO) cross sections.
As expected, the MC predicts that the sample passing the preselection is dominated by dilepton $\cPqt\cPaqt$.
The data yield is in good agreement with the prediction. We also quote the yields for
the LM0 and LM1 benchmark scenarios.

\begin{table}[htb]
\begin{center}
\caption{\label{tab:yields} Data yields and MC predictions after preselection, using the quoted NLO production cross sections $\sigma$.
The $\cPqt\cPaqt\to \ell^{+}\ell^{-}$ corrresponds  to dilepton $\cPqt\cPaqt$, including
$\cPqt \to \PW \to \Pgt \to \ell$; $\cPqt\cPaqt\to \mathrm{other}$ includes all other $\cPqt\cPaqt$ decay modes.
The samples of MC $\cPqt\cPaqt$, $\PW^{\pm}$ + jets, and single-top events were
generated with \MADGRAPH. The Drell--Yan sample (which includes events with
invariant masses as low as 10\GeVcc) was generated using a mixture of \MADGRAPH\ and
\PYTHIA. All other samples were generated with \PYTHIA.
The LM0 and LM1 benchmark scenarios are defined in the text. Uncertainties are statistical only.
}
\vspace{2 mm}
\begin{tabular}{lr|cccc}

\hline
              Sample                    &  $\sigma$ (pb)  &               $\Pe\Pe$   &            $\Pgm\Pgm$   &              $\Pe\Pgm$   &                 Total  \\
\hline
$\cPqt\cPaqt\rightarrow \ell^{+}\ell^{-}$  &  16.9           &  14.50 $\pm$ 0.24   &    17.52 $\pm$ 0.26   &    41.34 $\pm$ 0.40   &    73.36 $\pm$ 0.53  \\
$\cPqt\cPaqt\rightarrow \mathrm{other}$    &  140.6          &   0.49 $\pm$ 0.04   &     0.21 $\pm$ 0.03   &     1.02 $\pm$ 0.06   &     1.72 $\pm$ 0.08  \\
Drell--Yan                               &  18417          &   1.02 $\pm$ 0.21   &     1.16 $\pm$ 0.22   &     1.20 $\pm$ 0.22   &     3.38 $\pm$ 0.37  \\
    $\PW^{\pm}$ + jets                    &  28049          &   0.19 $\pm$ 0.13   &     0.00 $\pm$ 0.00   &     0.09 $\pm$ 0.09   &     0.28 $\pm$ 0.16  \\
            $\PW^+\PW^-$                    &  2.9            &   0.15 $\pm$ 0.01   &     0.16 $\pm$ 0.01   &     0.37 $\pm$ 0.02   &     0.68 $\pm$ 0.03  \\
        $\PW^{\pm}\cPZ$                      &  0.3            &   0.02 $\pm$ 0.00   &     0.02 $\pm$ 0.00   &     0.04 $\pm$ 0.00   &     0.09 $\pm$ 0.00  \\
            $\cPZ\cPZ$                        &  4.3            &   0.01 $\pm$ 0.00   &     0.02 $\pm$ 0.00   &     0.02 $\pm$ 0.00   &     0.05 $\pm$ 0.00  \\
          Single top                    &  33.0           &   0.46 $\pm$ 0.02   &     0.55 $\pm$ 0.02   &     1.24 $\pm$ 0.03   &     2.25 $\pm$ 0.04  \\
\hline					  					
         Total SM MC                    &                 &  16.85 $\pm$ 0.34   &    19.63 $\pm$ 0.34   &    45.33 $\pm$ 0.47   &    81.81 $\pm$ 0.67  \\
\hline					  					
                Data                    &                 &                15   &                  22   &                  45   &                  82  \\
\hline					  					
                 LM0                    &  52.9           &  10.67 $\pm$ 0.31   &    12.63 $\pm$ 0.34   &    17.81 $\pm$ 0.41   &    41.11 $\pm$ 0.62  \\
                 LM1                    &   6.7           &   2.35 $\pm$ 0.05   &     2.83 $\pm$ 0.06   &     1.51 $\pm$ 0.04   &     6.69 $\pm$ 0.09  \\

\hline

\end{tabular}
\end{center}
\end{table}

\begin{figure}[tbh]
\begin{center}
\includegraphics[width=1.0\linewidth]{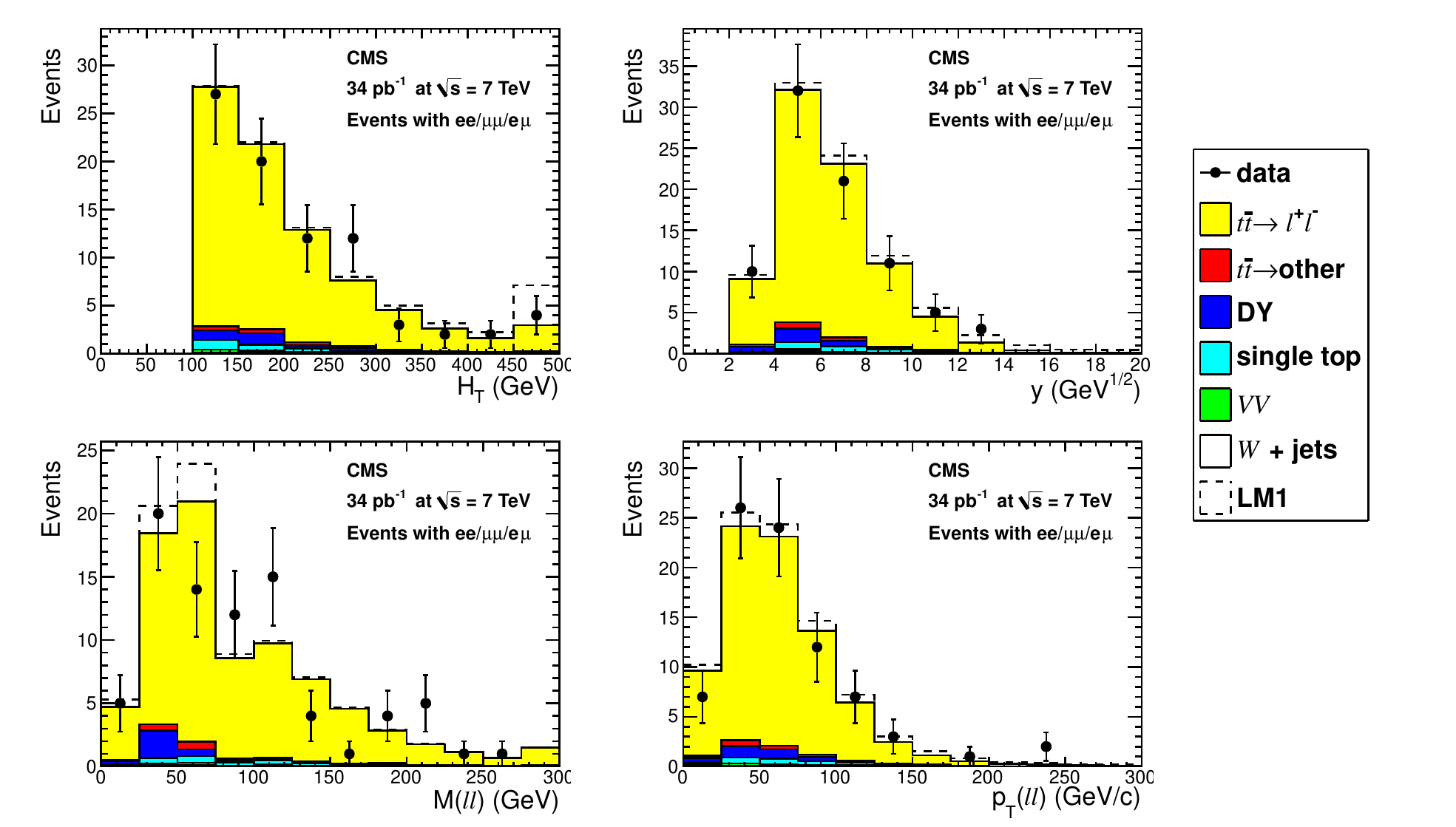}
\caption{\label{fig:bulk}\protect
Distributions of (top left) scalar sum of jet transverse energies (\HT),
(top right) $y\equiv\MET/\sqrt{\HT}$, 
(bottom left) dilepton invariant mass $M(\ell\ell)$,
and (bottom right) dilepton transverse momentum $\pt(\ell\ell)$ for SM MC and data after preselection.
The last bin contains the overflow.
Here $\cPqt\cPaqt\rightarrow \ell^{+}\ell^{-}$
corresponds  to dilepton $\cPqt\cPaqt$, including $\cPqt \to \PW \to \Pgt \to \ell$;
$\cPqt\cPaqt\rightarrow \mathrm{other}$ includes all other $\cPqt\cPaqt$ decay modes, and VV indicates the sum of
$\PW\PW$, $\PW\cPZ$, and $\cPZ\cPZ$. The MC distributions for the LM1 benchmark points are also shown.
}
\end{center}
\end{figure}

Figure~\ref{fig:bulk} compares several kinematic distributions
in data and SM MC for events passing the preselection.
As an illustration, we also show the MC distributions for the LM1 benchmark point.
We find that the SM MC reproduces
the properties of the bulk of dilepton $\cPqt\cPaqt$ events.
We therefore turn our attention to the tails of the \MET\ and \HT\
distributions of the $\cPqt\cPaqt$ sample.

To look for possible BSM contributions, we define a signal region that preserves about 1\%
of the dilepton $\cPqt\cPaqt$ events, by adding the following two requirements to the
preselection described above:

\begin{equation}
\label{eq:sigregion}
\HT > 300\GeV~\mathrm{and}~y > 8.5\GeV^{1/2},
\end{equation}

where $y \equiv \MET/\sqrt{\HT}$.
The requirement is on $y$ rather than \MET\
because the variables \HT\ and $y$ are found to be almost uncorrelated in dilepton $\cPqt\cPaqt$
MC, with a correlation coefficient of $\sim5$\%.
This facilitates the use of a background estimation method based on data,
as discussed in Section~\ref{sec:datadriven}.

The MC predicts 1.3 SM events, dominated by dilepton $\cPqt\cPaqt$, in
the signal region. The expectations for the LM0 and LM1 points are
8.6 and 3.6 events, respectively.

\section{Background Estimates from Data}
\label{sec:datadriven}
We have developed two independent methods to
estimate from data the background in the signal region.
The first method exploits the fact that
\HT\ and $y$ are nearly
uncorrelated for the $\cPqt\cPaqt$ background.
Four regions (A, B, C, and D) are defined in the
$y$ vs.\ \HT\ plane, as indicated in
Figure~\ref{fig:abcdData}, where region D is the signal region defined
in Eq.~\ref{eq:sigregion}.  In the absence of a signal, the yields in
the regions A, B, and C can be used to estimate
the yield in the signal region D as
$N_\textrm{D} = N_\textrm{A}\times{}N_\textrm{C}/N_\textrm{B}$; this method is referred to as the ``ABCD method".

The expected event yields
in the four regions for the SM MC, as well as the background
prediction $N_\textrm{A}\times{}N_\textrm{C}/N_\textrm{B}$, are given in Table~\ref{tab:datayield}.
We observe good agreement between the total SM MC predicted and observed yields.
A 20\% systematic uncertainty is assigned to
the predicted yield of the ABCD method to take into account uncertainties from contributions of
backgrounds other than dilepton $\cPqt\cPaqt$ (16\%), finite MC statistics in the closure test (8\%),
and variation of the boundaries between the ABCD regions based on the uncertainty in the
hadronic energy scale (8\%).

The second  background estimate, henceforth referred to as the dilepton transverse momentum ($\pt(\ell\ell)$) method,
is  based on the  idea~\cite{ref:victory} that  in dilepton  $\cPqt\cPaqt$  events the
\pt\  distributions of  the charged  leptons and  neutrinos  from $\PW$
decays are  related, because of the  common boosts from  the top  and $\PW$
decays.  This relation  is governed by the polarization  of the $\PW$'s,
which         is         well         understood        in         top
decays in the SM~\cite{Wpolarization,Wpolarization2}   and   can  therefore   be
reliably  accounted   for.   We then  use   the  observed
$\pt(\ell\ell)$ distribution to  model the $\pt(\nu\nu)$ distribution,
which is  identified with \MET.  Thus,  we use the  number of observed
events  with $\HT >  300\GeV$  and $\pt(\ell\ell)/\sqrt{\HT}  >8.5\GeV^{1/2}$
to predict the  number of  background events  with $\HT >  300\GeV$ and
$y = \MET/\sqrt{\HT}>8.5\GeV^{1/2}$.   In  practice, two corrections must be applied
to this prediction, as described below.

The first correction  accounts for the $\MET >  50\GeV$ requirement in the
preselection, which is needed to  reduce the DY background.  We
rescale  the  prediction by  a  factor equal  to  the  inverse of  the
fraction  of  events  passing  the preselection which  also  satisfy  the
requirement  $\pt(\ell\ell) >  50\GeVc$.  This  correction  factor is
determined  from  MC  and  is  $K_{50}=1.5$.   The  second  correction
($K_C$) is  associated with the  known polarization  of the  $W$, which
introduces a difference  between the $\pt(\ell\ell)$ and $\pt(\nu\nu)$
distributions. The correction $K_C$ also takes into account detector effects such as the hadronic energy
scale and  resolution which affect  the \MET\ but  not $\pt(\ell\ell)$.
The  total correction factor  is $K_{50}  \times K_C  = 2.1  \pm 0.6$,
where  the uncertainty  is dominated  by  the 5\%  uncertainty in  the
hadronic energy scale~\cite{ref:jes}.

All background estimation methods based on data are in principle subject to signal contamination
in the control regions, which tends to decrease the significance of a signal
which may be present in the data by increasing the background prediction.
In general, it is difficult to quantify these effects because we
do not know what signal may be present in the data.  Having two
independent methods (in addition to expectations from MC)
adds redundancy because signal contamination can have different effects
in the different control regions for the two methods.
For example, in the extreme case of a
BSM signal with identical distributions of $\pt(\ell \ell)$ and \MET, an excess of events might be seen
in the ABCD method but not in the $\pt(\ell \ell)$ method.

Backgrounds in which one or both leptons do not originate from electroweak decays (non-$\PW/\cPZ$ leptons)
are assessed using the method of Ref.~\cite{ref:top}.
A non-$\PW/\cPZ$ lepton is a lepton candidate originating from within a jet,
such as a lepton from semileptonic $\cPqb$ or $\cPqc$ decays, a muon decay-in-flight, a pion misidentified
as an electron, or an unidentified photon conversion.
Estimates of the contributions to the signal region from pure
multijet QCD, with two non-$\PW/\cPZ$ leptons, and in $\PW+\mathrm{jets}$,
with one non-$\PW/\cPZ$ lepton in addition to the lepton from the decay of the $\PW$, are
derived separately. We find $0.00^{+0.04}_{-0.00}$ and $0.0^{+0.4}_{-0.0}$ for the multijet
QCD and $\PW$+jets contributions respectively,
and thus consider these backgrounds to be negligible.

Backgrounds from DY and from processes with two vector bosons and single top
are negligible compared to dilepton $\cPqt\cPaqt$.

\section{Results}
\label{sec:results}

We find one event in the signal region D. The event is in the $\Pe\Pgm$ channel and contains 3 jets.
The SM MC expectation is 1.3 events.

\begin{figure}[tbh]
\begin{center}
\includegraphics[width=0.75\linewidth]{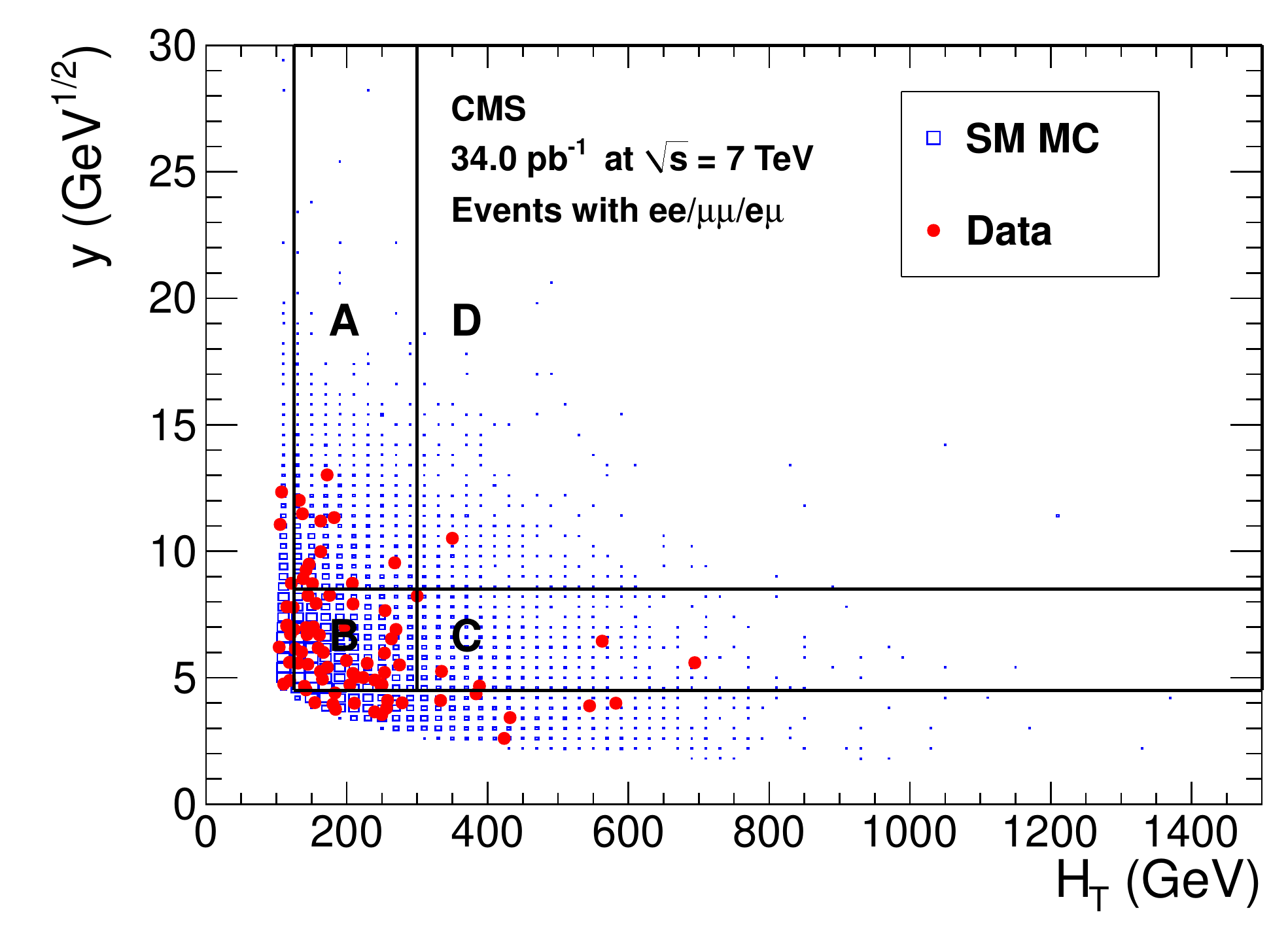}
\caption{\label{fig:abcdData}\protect Distributions of $y$ vs.\ \HT\
for SM MC (2-dimensional histogram) and data (scatter plot).  Here  our
choice of the ABCD regions is also shown.}
\end{center}
\end{figure}

Table~\ref{tab:datayield} summarizes the event yields obtained for each of the four ABCD
regions in the data and in the  MC samples.
The prediction of the ABCD method is given by
$N_\textrm{A}\times{}N_\textrm{C}/N_\textrm{B} = 1.3 \pm 0.8~(\textrm{stat.}) \pm 0.3~(\textrm{syst.})$ events.
The data, together with SM expectations, are presented in Figure~\ref{fig:abcdData}.

\begin{table}[hbt]
\begin{center}
\caption{\label{tab:datayield} Data yields in the four
regions of Figure~\ref{fig:abcdData}, as well as the predicted yield in region D given
by $N_\textrm{A}\times{}N_\textrm{C}/N_\textrm{B}$. The SM and BSM MC expectations are also shown.
The quoted uncertainties are statistical only.}
\vspace{2 mm}
\begin{tabular}{l||c|c|c|c||c}
\hline
              Sample                    &        $N_\text{A}$  &         $N_\text{B}$  &         $N_\text{C}$  &          $N_\text{D}$ &$N_\textrm{A}\times{}N_\textrm{C}/N_\textrm{B}$\\
\hline
$\cPqt\cPaqt\rightarrow \ell^{+}\ell^{-}$  &  8.44  $\pm$  0.18   &  32.83  $\pm$  0.35   &   4.78  $\pm$  0.14   &   1.07  $\pm$  0.06   &   1.23  $\pm$  0.05  \\
$\cPqt\cPaqt\rightarrow \mathrm{other}$    &  0.12  $\pm$  0.02   &   0.78  $\pm$  0.05   &   0.16  $\pm$  0.02   &   0.02  $\pm$  0.01   &   0.02  $\pm$  0.01  \\
Drell--Yan                               &  0.17  $\pm$  0.08   &   1.18  $\pm$  0.22   &   0.04  $\pm$  0.04   &   0.12  $\pm$  0.07   &   0.01  $\pm$  0.01  \\
    $\PW^{\pm}$ + jets                    &  0.00  $\pm$  0.00   &   0.09  $\pm$  0.09   &   0.00  $\pm$  0.00   &   0.00  $\pm$  0.00   &   0.00  $\pm$  0.00  \\
            $\PW^+\PW^-$                    &  0.11  $\pm$  0.01   &   0.29  $\pm$  0.02   &   0.02  $\pm$  0.01   &   0.03  $\pm$  0.01   &   0.01  $\pm$  0.00  \\
        $\PW^{\pm}Z$                      &  0.01  $\pm$  0.00   &   0.04  $\pm$  0.00   &   0.00  $\pm$  0.00   &   0.00  $\pm$  0.00   &   0.00  $\pm$  0.00  \\
            $\cPZ\cPZ$                        &  0.01  $\pm$  0.00   &   0.02  $\pm$  0.00   &   0.00  $\pm$  0.00   &   0.00  $\pm$  0.00   &   0.00  $\pm$  0.00  \\
          Single top                    &  0.29  $\pm$  0.01   &   1.04  $\pm$  0.03   &   0.04  $\pm$  0.01   &   0.01  $\pm$  0.00   &   0.01  $\pm$  0.00  \\
\hline
         Total SM MC                    &  9.14  $\pm$  0.20   &  36.26  $\pm$  0.43   &   5.05  $\pm$  0.14   &   1.27  $\pm$  0.10   &   1.27  $\pm$  0.05  \\
\hline
                Data                    &                 12   &                  37   &                   4   &                   1   &   1.30  $\pm$  0.78  \\
\hline
\hline
                 LM0                    &   4.04  $\pm$  0.19  &   4.45  $\pm$  0.20   &  13.92  $\pm$  0.36   &   8.63  $\pm$  0.27   &  12.63  $\pm$  0.88  \\
                 LM1                    &   0.52  $\pm$  0.02  &   0.26  $\pm$  0.02   &   1.64  $\pm$  0.04   &   3.56  $\pm$  0.06   &   3.33  $\pm$  0.27  \\
\hline
\end{tabular}
\end{center}
\end{table}

\begin{figure}[hbt]
\begin{center}
\includegraphics[width=0.49\linewidth]{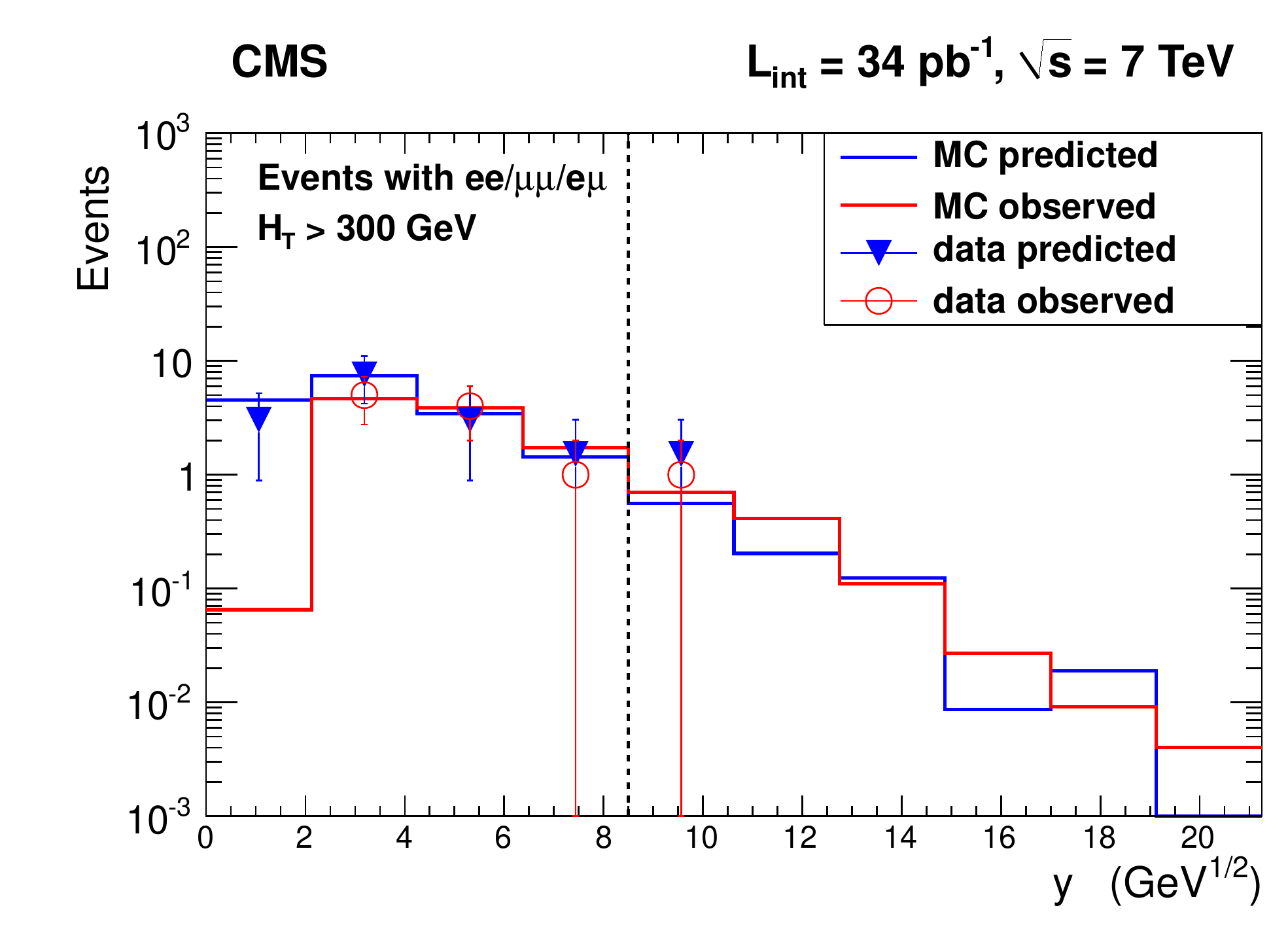}
\includegraphics[width=0.49\linewidth]{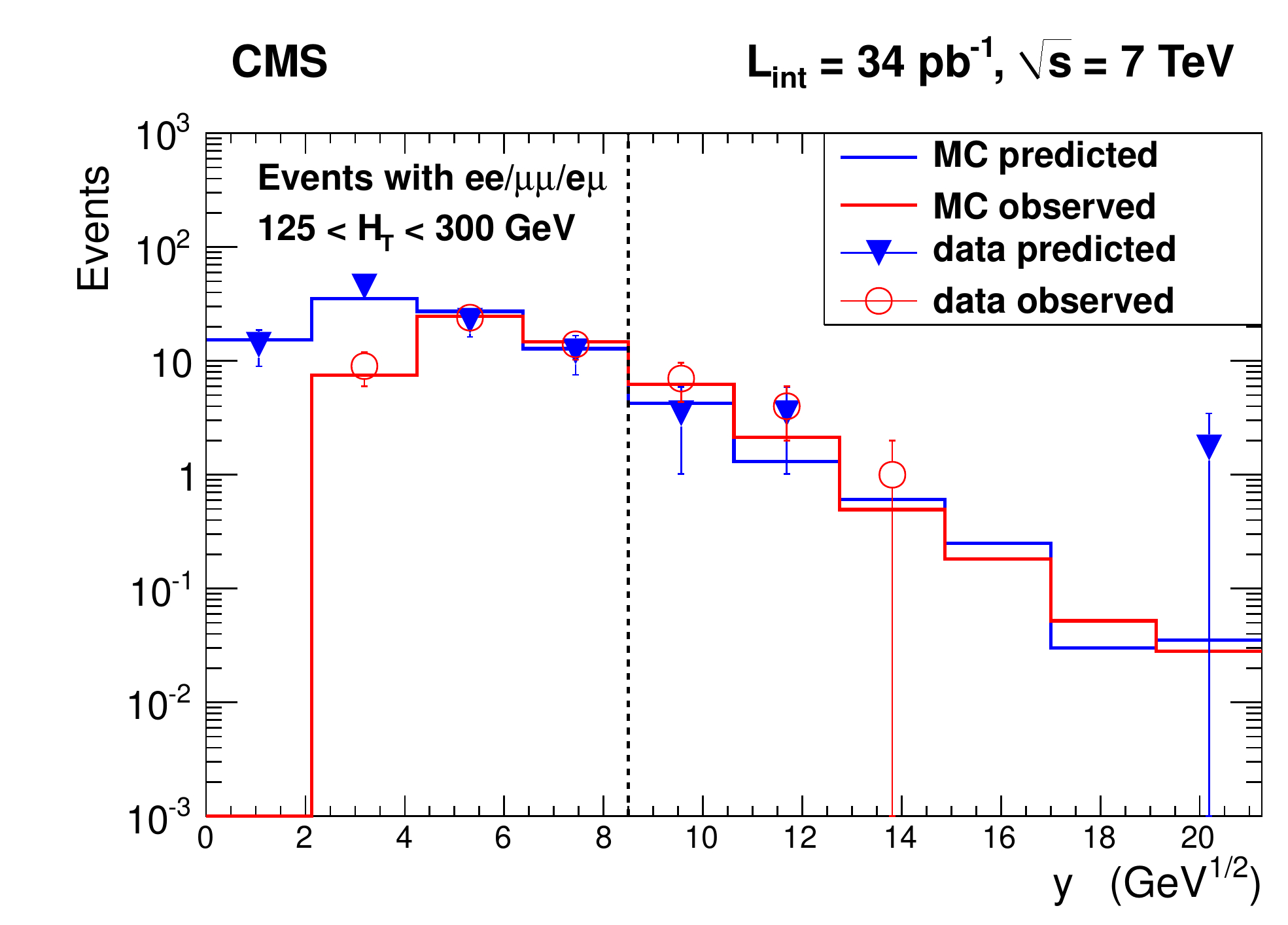}
\caption{\label{fig:victory}\protect Distributions of
$y$ (observed) and ${\pt(\ell\ell)}/\sqrt{\HT}$ scaled by the correction
factor $K_{50}$ (predicted)
for (left) the signal region and (right) the control region A, for both MC and data.
The vertical dashed line indicates the search region defined by $y>8.5\GeV^{1/2}$.
The deficit at low $y$ is due to the $\MET > 50\GeV$ preselection requirement.}
\end{center}
\end{figure}

The ABCD prediction is then compared with that of the $\pt(\ell \ell)$ method.
We find 1 event passing the requirements $\HT > 300\GeV$ and
$\pt(\ell\ell)/\sqrt{\HT} > 8.5\GeV^{1/2}$. This leads to a predicted
background of $2.1 \pm 2.1~({\rm stat.}) \pm 0.6~({\rm syst.})$ after applying
the correction factor $K_{50} \times K_C = 2.1 \pm 0.6$,
as shown in Figure~\ref{fig:victory} (left).

As a validation of the $\pt(\ell\ell)$ method in a region with higher statistics,
we also apply the $\pt(\ell\ell)$ method in control region A by restricting
\HT\ to be in the range 125--300~\GeV. Here the prediction is $9.0 \pm 6.0~(\textrm{stat.})$
background events, in good agreement with the observed yield of 12 events, as shown
in Figure~\ref{fig:victory} (right).

In summary, for the signal region defined as $\HT>300\GeV$ and $y > 8.5\GeV^{1/2}$:
we observe one event in the data,
SM MC predicts 1.3 events,
the ABCD method predicts $1.3 \pm 0.8~({\rm stat.}) \pm 0.3~({\rm syst.})$ events,
and the $\pt(\ell\ell)$ method predicts $2.1 \pm 2.1~({\rm stat.}) \pm 0.6~({\rm syst.})$ events.

All three background predictions are consistent within their uncertainties.
We thus take as our best estimate of the SM yield in
the signal region the error-weighted average of the two background estimates based on data and find
a number of predicted background events $N_\textrm{BG}=1.4 \pm 0.8$, in good agreement with the
observed signal yield. We therefore conclude that no evidence for a non-SM contribution
to the signal region is observed.

\section{Acceptance and Efficiency Systematic Uncertainties}
\label{sec:systematics}

The acceptance and efficiency, as well as the systematic uncertainties in these quantities,
depend on the signal model.
For some of the individual uncertainties, it is reasonable to quote values
based on SM control samples with kinematic properties similar to the SUSY benchmark models.
For others that depend strongly on the kinematic properties of the event, the systematic
uncertainties must be quoted model by model.

The systematic uncertainty in the lepton acceptance consists
of two parts: the trigger efficiency uncertainty and the
identification and isolation uncertainty. The trigger efficiency
for two leptons of $\pt>10\GeVc$, with one lepton of
$\pt>20\GeVc$ is close to 100\%.
We estimate the efficiency uncertainty to be a few percent,
mostly in the low \pt\ region, using samples of $\cPZ \to \ell\ell$.
For dilepton $\cPqt\cPaqt$, LM0, and LM1,
the trigger efficiency uncertainties are found to be less than 1\%.
We verify that the MC reproduces the lepton identification and isolation efficiencies in data using
samples of $\cPZ \to \ell\ell$; the data and MC efficiencies are found to be consistent within 2\%.

Another significant source of systematic uncertainty is
associated with the jet and $\MET$ energy scale.  The impact
of this uncertainty is final-state dependent.  Final
states characterized by very large hadronic activity and \MET\ are
less sensitive than final states where the \MET\ and \HT\
are typically close to the minimum requirements applied to these quantities.  To be more quantitative,
we have used the method of Ref.~\cite{ref:top} to evaluate
the systematic uncertainties in the acceptance for $\cPqt\cPaqt$
and for the two benchmark SUSY points using a 5\% uncertainty in the hadronic
energy scale~\cite{ref:jes}.
For $\cPqt\cPaqt$ the uncertainty is 27\%; for LM0 and LM1 the
uncertainties are 14\% and 6\%, respectively.

The uncertainty in the integrated luminosity is 11\%~\cite{ref:lumi}.

\section{Same-flavour Dilepton Search}
\label{sec:HT}

The result of Section~\ref{sec:results} is cross-checked in a similar kinematic region with an
independent
search relying on a different trigger path, different methods for ``physics object'' reconstruction, and a
different background estimation method.
This search is directed at BSM scenarios in which decay chains of a pair of new heavy particles
produce an excess of same-flavour ($\eepm$ and $\mmpm$) events over opposite-flavour ($\empm$) events.
For example, in the context of the CMSSM, this excess may be caused by decays of neutralinos and \cPZ\ bosons to same-flavour lepton pairs.
For the benchmark scenario LM0 (LM1), the fraction of same-flavour events in the signal region discussed
below is 0.67 (0.86).

The dominant background in this search is also dilepton $\cPqt\cPaqt$, for which such an excess does not exist
because the flavours of the two leptons are uncorrelated.
Therefore, the rate of $\cPqt\cPaqt$ decays with two  same-flavour leptons
may be estimated from the number of opposite-flavour events,
after correcting for the
ratio of muon to electron selection efficiencies, $r_{\Pgm\Pe}$.
This method actually estimates the contribution of any uncorrelated pair of leptons, including
e.g.\ $\cPZ\to\tau\tau$ events where the two $\tau$ leptons decay leptonically.
This method will also subtract any BSM signal producing lepton pairs of uncorrelated flavour.

Events with two leptons with $\pt>10\GeVc$ are selected. Because the lepton triggers are not fully
efficient for events with two leptons of $\pt>10\GeVc$,
the data sample for this analysis is selected with hadronic triggers based on the
scalar sum of the transverse energies of all jets reconstructed from calorimeter signals with $\pt>20\GeVc$.
The event is required to pass at least one of a set of hadronic triggers with transverse energy thresholds
ranging from 100 to 150~GeV. The efficiency of this set of triggers with respect to the analysis selection is
greater than 99\%.
In addition to the trigger, we require $\HT>350\GeV$,
where \HT\ in this analysis is defined as the scalar sum of
the transverse energies of all selected jets with $\pt>30\GeVc$
and within an increased pseudorapidity range $|\eta|<3$, in line with the trigger requirement.
The jets, \MET, and leptons are reconstructed with the Particle Flow technique~\cite{CMS-PAS-PFT-10-002}.
The resulting performance of the selection of leptons and jets does not differ
significantly from the selection discussed  in Section~\ref{sec:eventSel}.

The signal region is defined by additionally requiring $\MET>150\GeV$.
This signal region is chosen such that approximately one SM event is expected in our
current data sample.

The  lepton   selection  efficiencies  are  measured   using  the  $Z$
resonance.    As   discussed   in  Section~\ref{sec:systematics},   these
efficiencies are  known with a  systematic uncertainty of  $2\%$.  The
selection efficiencies of isolated leptons are different in the $t\bar{t}$
and  $\cPZ+\textrm{jets}$  samples.   The   ratio  of  muon  to  electron
efficiencies $r_{\Pgm\Pe}$,  however, is found  to differ by  less than
5\%  in the MC simulations,  and  a corresponding  systematic
uncertainty is assigned to this ratio. This procedure gives $r_{\Pgm\Pe} = 1.07 \pm
0.06$.

The $\PW+\textrm{jets}$ and QCD multijet  contributions, where at least one
of the two leptons is a  secondary lepton from a heavy flavour decay or
a jet misidentified as a lepton (non-$\PW/\cPZ$ leptons) are estimated from
a  fit to the  lepton isolation distribution,  after relaxing
the  isolation requirement on  the leptons.
Contributions from
other SM backgrounds,  such as DY or processes  with two gauge bosons,
are strongly suppressed  by the \MET\ requirement and  are expected to
be negligible.

We first estimate the number of SM events in a \ttbar-dominated region
with $100 < \HT < 350\GeV$ and $\MET>80\GeV$. In order to cope with the lower \HT\ requirement,
we use the same high-\pt lepton trigger sample as described in Section~\ref{sec:eventSel}.
In this region we observe $26$ opposite-flavour candidates and predict $1.0\pm0.5$
non-$W/Z$ lepton events from the fit to the lepton isolation distribution. This results in an
estimate of $25.0 \pm 5.0$ \ttbar\ events in the $\Pe\Pgm$ channel. Using the efficiency
ratio  $r_{\Pgm\Pe}$  this estimate is then converted into a prediction for the number
of same-flavour events in the $\Pe\Pe$ and $\Pgm\Pgm$ channels.

\begin{table}[hbt]
\begin{center}
\caption{\label{tab:CRresults}
Number of predicted and observed $\Pe\Pe$ and $\Pgm\Pgm$
events in the control region, defined as $100 < \HT < 350\GeV$ and $\MET > 80\GeV$.
``SM MC'' indicates the sum of all MC samples ($\cPqt\cPaqt$, DY, $\PW+\textrm{jets}$, and $\PW\PW/\PW\cPZ/\cPZ\cPZ$)
and includes statistical uncertainties only.
}
\vspace{2 mm}
\begin{tabular}{l|cc}
\hline
                                 & \multicolumn{2}{c}{Control region}               \\
\hline
Process                          & $ee$          & $\mu\mu$        \\
\hline
$\cPqt\cPaqt$ predicted from $\Pe\Pgm$ & $11.7\pm 2.4$ & $13.4\pm 2.8$   \\
Non-$\PW/\cPZ$ leptons                & $0.5\pm 0.3$  & $0.4\pm0.2$ \\
\hline
Total predicted                  & $12.2\pm 2.4$ & $13.8 \pm 2.8$  \\
\hline\hline
Total observed                   & $10$          & $15$          \\
\hline \hline
SM MC                            & $8.4\pm 0.2$  & $10.5 \pm 0.3$    \\

\hline
\end{tabular}
\end{center}
\end{table}

Table~\ref{tab:CRresults} shows the number of expected SM background same-flavour events in the control region for the MC,
as well as the prediction from the background estimation techniques based on data. There are a total of 25 same-flavour
events, in good agreement with the prediction of $25.9 \pm 5.2$ events.
We thus proceed to the signal region selection.

The SM background predictions in the signal region from the opposite-flavour and non-$W/Z$ lepton methods
are summarized in Table~\ref{tab:results}.
We find one event in the signal region in the $\Pe\Pgm$ channel with a prediction of non-$\PW/\cPZ$ leptons
of $0.1\pm0.1$, and thus predict $0.9 {}_{-0.8}^{+2.2}$ same-flavour events using Poisson statistical uncertainties.
In the data we find no same-flavour events, in agreement with the prediction, in contrast with $7.3\pm1.6$
and $3.6\pm0.7$ expected events for the benchmark points LM0 and LM1, respectively.
The predicted background from non-$\PW/\cPZ$ leptons is negligible.

\begin{table}[hbt]
\begin{center}
\caption{\label{tab:results}
Number of predicted and observed events in the signal region, defined as $\HT > 350\GeV$ and $\MET> 150\GeV$.
``SM MC'' indicates the sum of all MC samples ($\cPqt\cPaqt$, DY, $\PW+\textrm{jets}$, and $\PW\PW/\PW\cPZ/\cPZ\cPZ$)
and includes statistical uncertainties only.
}
\vspace{2 mm}
\begin{tabular}{l|cc}
\hline
                                    &   \multicolumn{2}{c}{Signal region}          \\
\hline
Process                             & $\Pe\Pe$                   & $\Pgm\Pgm$  \\
\hline
$\cPqt\cPaqt$ predicted from $\Pe\Pgm$    & $0.4 {}_{-0.4}^{+1.0}$ & $0.5 {}_{-0.4}^{+1.2}$  \\
Non-$\PW/\cPZ$                           & 0                      & 0                        \\
\hline
Total predicted                     & $0.4 {}_{-0.4}^{+1.0}$ & $0.5 {}_{-0.4}^{+1.2}$   \\
\hline\hline
Total observed                      & $0$                    & $0$  \\
\hline \hline
SM MC                               & $0.38\pm 0.08$         & $0.56 \pm 0.07$ \\
LM0                                 & $3.4\pm0.2$            & $3.9\pm0.2$  \\
LM1                                 & $1.6\pm0.1$            & $2.0\pm0.1$  \\

\hline
\end{tabular}
\end{center}
\end{table}

Table~\ref{tab:results} demonstrates the sensitivity of this approach.
We observe comparable yields of the same benchmark points as for the high-\pt\
lepton trigger search, where 35--60\% of the events are common to both
searches for LM0 and LM1.
Either approach would have given an excess in the presence of a signal.

\section{Limits on New Physics}
\label{sec:limit}

The three background predictions for the high-\pt\ lepton trigger search
discussed in Section~\ref{sec:results} are in good agreement with each other and
with the observation of one event in the signal region. 
A Bayesian 95\%~confidence level (CL) upper limit~\cite{ref:cl95cms}  on the number of
non-SM events in the signal region is determined to be 4.0,
using a background prediction of $N_\textrm{BG} = 1.4 \pm 0.8$
events and a log-normal model of nuisance parameter integration.
The upper limit is not very sensitive to $N_\textrm{BG}$ and its uncertainty.
This generic upper limit is not corrected for the possibility
of signal contamination in the control regions. This is justified because
the two independent background estimation methods based on data agree
and are also consistent with the SM MC prediction.
Moreover,  no evidence for non-SM contributions in
the control regions is observed (Table~\ref{tab:datayield} and Figure~\ref{fig:victory}). 
This bound rules out the benchmark SUSY scenario LM0, for which the
number of expected signal events is $8.6 \pm 1.6$, while the LM1 scenario predicts
$3.6 \pm 0.5$ events.
The uncertainties in the LM0 and LM1 event yields arise from energy scale, luminosity, 
and lepton efficiency, as discussed in Section~\ref{sec:systematics}.

For the same-flavour search using hadronic activity triggers discussed in Section~\ref{sec:HT}, 
no same-flavour events are observed and the corresponding Bayesian 95\% CL upper limit on the 
non-SM yield is 3.0 events. 
This bound rules out the benchmark SUSY scenarios LM0 and LM1, for which the
numbers of expected signal events are $7.3 \pm 1.6$ and $3.6 \pm 0.7$, respectively.

\begin{figure}[tbh]
\begin{center}
\includegraphics[width=1\linewidth]{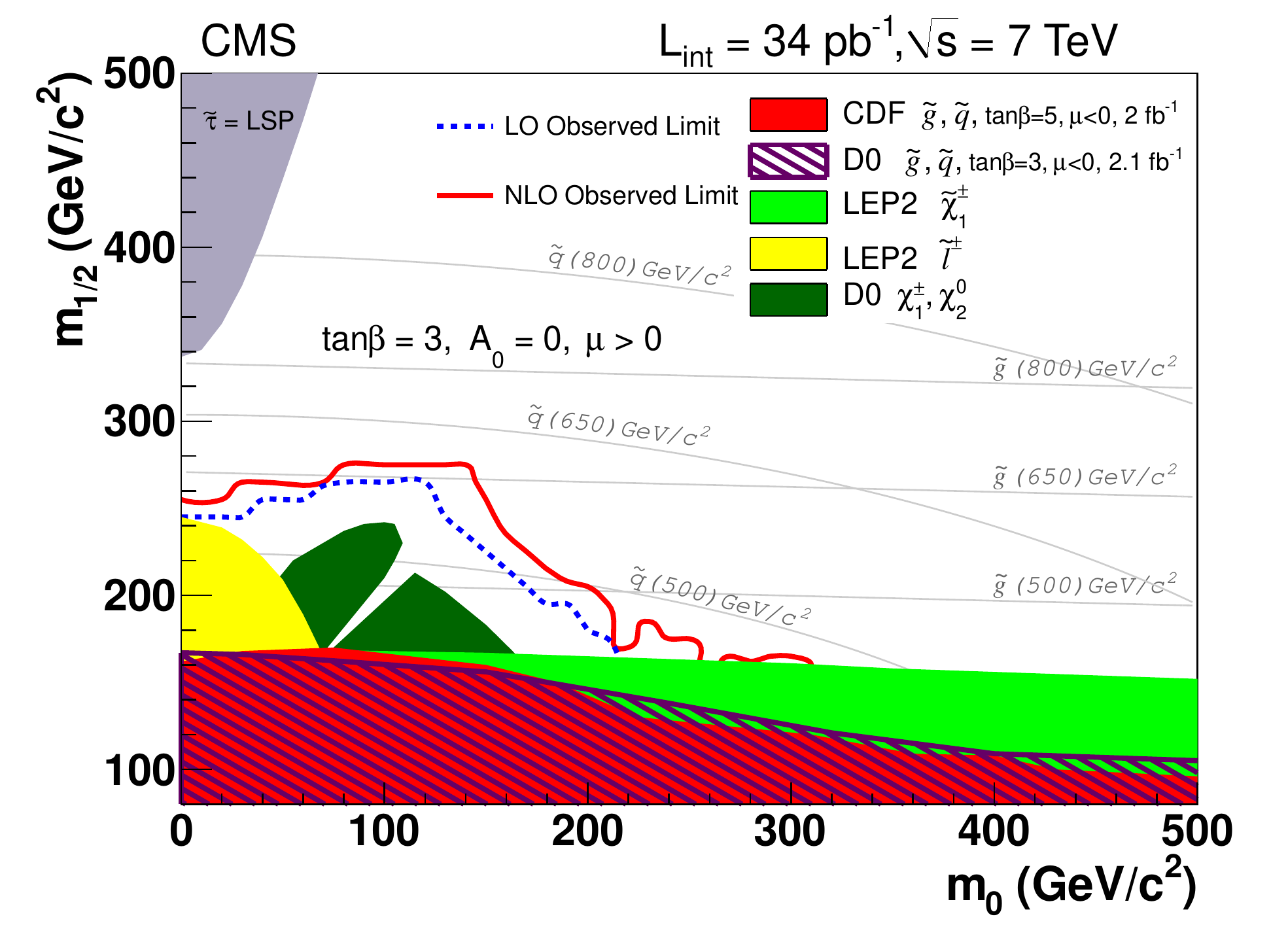}
\caption{\label{fig:msugra}\protect 
The observed 95\% CL exclusion contour at NLO (solid red line) and LO (dashed blue line) 
in the CMSSM $(m_0,m_{1/2})$ plane for  $\tan\beta=3$, $A_0 = 0$ and $\mu > 0$. 
The area below the curve is excluded by this measurement. Exclusion limits obtained from 
previous experiments are presented as filled areas in the plot. Thin grey lines correspond to 
constant squark and gluino masses.
}
\end{center}
\end{figure}

We also quote the result more generally in the context of the CMSSM.
The Bayesian 95\% CL limit  in the  $(m_0,m_{1/2})$ plane,  for $\tan\beta=3$,
$A_0 = 0$ and $\mu > 0$ is shown in Figure~\ref{fig:msugra}. 
The high-\pt\ lepton and hadronic trigger searches have similar sensitivity to the CMSSM; 
here we choose to show results based on the high-\pt\ lepton trigger search. The
SUSY particle  spectrum is calculated using  SoftSUSY~\cite{Allanach:2002uq}, and the
signal  events  are  generated  at  leading  order  (LO)  with  \PYTHIA6.4.22.
NLO  cross sections,  obtained  with the
program  Prospino~\cite{Beenakker:1997kx},  are used  to  calculate  the observed
exclusion  contour.  
At each point in the  $(m_0,m_{1/2})$ plane, the acceptance uncertainty is calculated by
summing in quadrature the uncertainties from jet and \MET\ energy scale using the
procedure discussed in Section~\ref{sec:systematics}, the uncertainty in the 
NLO cross section due to the choice of factorization and renormalization scale, 
and the uncertainty from the parton distribution function (PDF) for CTEQ6.6~\cite{Nadolsky:2008fk},
estimated from  the  envelope  provided  by  the  CTEQ6.6  error sets.
The luminosity uncertainty and dilepton
selection efficiency uncertainty are also included, giving a total relative acceptance uncertainty which varies
in the range 0.2--0.3.
A point is considered to be excluded if the NLO yield exceeds the 95\% CL
Bayesian upper limit calculated with this acceptance uncertainty, using
a log-normal model for the nuisance parameter integration.
The limit curves do not include the effect of signal
contamination in the control regions.  We have verified that this
has a negligible impact on the excluded regions in Figure~\ref{fig:msugra}.

The excluded regions  for the CDF
search for  jets + missing  energy final states~\cite{PhysRevLett.102.121801} were
obtained for $\tan\beta=5$, while those from D0~\cite{Abazov2008449} were obtained for 
$\tan\beta=3$, each with  approximately  2~fb$^{-1}$ of  data and for $\mu < 0$. 
The  LEP-excluded
regions  are based  on searches  for  sleptons and  charginos~\cite{LEPSUSY}.  
The D0 exclusion limit, valid for $\tan\beta=3$  and obtained from
a  search  for  associated  production   of  charginos $\chi_{1}^{\pm}$ and
neutralinos $\chi_2^0$ in  trilepton final states~\cite{Abazov200934}, is also
included  in Figure~\ref{fig:msugra}. In  contrast to  the other  limits  presented in
Figure~\ref{fig:msugra},  the results of our search and of the  trilepton search are  strongly dependent on
the choice  of $\tan\beta$ and  they reach the  highest sensitivity  in the
CMSSM for $\tan\beta$ values below 10.

\section{Additional Information for Model Testing}
\label{sec:outreach}
Other models of new physics in the dilepton final state can be confronted in an approximate way by simple
generator-level studies that compare the expected number of events in 34\pbinv\
with the upper limits from Section~\ref{sec:limit}.
The key ingredients of such studies are the kinematic requirements described
in this paper, the lepton efficiencies, and the detector responses for \HT, $y$, and \MET.
The muon identification efficiency is $\approx 95\%$;
the electron identification efficiency varies approximately linearly from $\approx$ 63\% at
$\pt = 10\GeVc$ to 91\% for $\pt > 30\GeVc$.
The lepton isolation efficiency depends on the lepton momentum, as well as on the jet activity in the
event.
In $\cPqt\cPaqt$ events, it varies approximately linearly from $\approx 83\%$ (muons)
and $\approx 89\%$ (electrons) at $\pt=10\GeVc$ to $\approx 95\%$ for $\pt>60\GeVc$.
In LM0 events, this efficiency is decreased by $\approx 5$--10\% over the whole momentum spectrum.
Electrons and muons from LM1 events have the same isolation efficiency as in $\cPqt\cPaqt$ events
at low \pt\ and $\approx 90$\% efficiency for $\pt>60\GeVc$.
The average detector responses (the reconstructed quantity divided by the generated quantity)
for \HT, $y$ and \MET\ are consistent with 1 within the 5\% jet energy scale uncertainty.
The experimental resolutions on these quantities are 10\%, 14\% and 16\%, respectively.

\section{Summary}
\label{sec:conclusion}

We have presented a search for BSM physics in the opposite-sign dilepton final state using 
a data sample of proton-proton collisions at 7~TeV centre-of-mass energy corresponding to an integrated
luminosity of \lumifinal, recorded by the CMS detector in 2010.
The search focused on dilepton events with large missing transverse energy and significant hadronic activity,
motivated by many models of BSM physics, such as supersymmetric models.
Good agreement with standard model predictions was found, both in terms of event yields and shapes of 
various relevant kinematic distributions. In the absence of evidence for BSM physics, 
we have set upper limits on the non-SM contributions to the signal regions. The result was interpreted
in the context of the CMSSM parameter space and the excluded region was found to exceed those set by
previous searches at the Tevatron and LEP experiments. 
Information on the acceptance and efficiency of the search was also provided to 
allow testing the exclusion of specific models of BSM physics.

\section*{Acknowledgments}

We wish to congratulate our colleagues in the CERN accelerator departments for the excellent performance of the LHC machine. We thank the technical and administrative staff at CERN and other CMS institutes, and acknowledge support from: FMSR (Austria); FNRS and FWO (Belgium); CNPq, CAPES, FAPERJ, and FAPESP (Brazil); MES (Bulgaria); CERN; CAS, MoST, and NSFC (China); COLCIENCIAS (Colombia); MSES (Croatia); RPF (Cyprus); Academy of Sciences and NICPB (Estonia); Academy of Finland, ME, and HIP (Finland); CEA and CNRS/IN2P3 (France); BMBF, DFG, and HGF (Germany); GSRT (Greece); OTKA and NKTH (Hungary); DAE and DST (India); IPM (Iran); SFI (Ireland); INFN (Italy); NRF and WCU (Korea); LAS (Lithuania); CINVESTAV, CONACYT, SEP, and UASLP-FAI (Mexico); PAEC (Pakistan); SCSR (Poland); FCT (Portugal); JINR (Armenia, Belarus, Georgia, Ukraine, Uzbekistan); MST and MAE (Russia); MSTD (Serbia); MICINN and CPAN (Spain); Swiss Funding Agencies (Switzerland); NSC (Taipei); TUBITAK and TAEK (Turkey); STFC (United Kingdom); DOE and NSF (USA).

\bibliography{auto_generated}   

\cleardoublepage\appendix\section{The CMS Collaboration \label{app:collab}}\begin{sloppypar}\hyphenpenalty=5000\widowpenalty=500\clubpenalty=5000\textbf{Yerevan Physics Institute,  Yerevan,  Armenia}\\*[0pt]
S.~Chatrchyan, V.~Khachatryan, A.M.~Sirunyan, A.~Tumasyan
\vskip\cmsinstskip
\textbf{Institut f\"{u}r Hochenergiephysik der OeAW,  Wien,  Austria}\\*[0pt]
W.~Adam, T.~Bergauer, M.~Dragicevic, J.~Er\"{o}, C.~Fabjan, M.~Friedl, R.~Fr\"{u}hwirth, V.M.~Ghete, J.~Hammer\cmsAuthorMark{1}, S.~H\"{a}nsel, M.~Hoch, N.~H\"{o}rmann, J.~Hrubec, M.~Jeitler, G.~Kasieczka, W.~Kiesenhofer, M.~Krammer, D.~Liko, I.~Mikulec, M.~Pernicka, H.~Rohringer, R.~Sch\"{o}fbeck, J.~Strauss, F.~Teischinger, P.~Wagner, W.~Waltenberger, G.~Walzel, E.~Widl, C.-E.~Wulz
% ...
\vskip\cmsinstskip
% ...
1:~~Also at CERN, European Organization for Nuclear Research, Geneva, Switzerland\\
% ...

\end{sloppypar}
\end{document}